\documentclass[twocolumn]{jpsj2}
\usepackage{graphicx}

\title{Bosons in optical lattices - from the Mott transition to the Tonks-Girardeau gas} 

\author{S. Wessel$^{(1,2)}$, F.  Alet$^{(2,3,4)}$, S. Trebst$^{(2,3)}$, D. Leumann$^{(2)}$,
M. Troyer$^{(2,3)}$, G. George Batrouni$^{(5)}$}

\inst{
$^{(1)}$Institut f\"ur Theoretische Physik III, Universit\"at
Stuttgart, 70550 Stuttgart, Germany \\ $^{(2)}$Theoretische Physik,
ETH Z\"urich, CH-8093 Z\"urich, Switzerland \\ $^{(3)}$Computational
Laboratory, ETH Z\"urich, CH-8092 Z\"urich, Switzerland \\
$^{(4)}$Service de Physique Th{\'e}orique, CEA Saclay, 91191 Gif sur
Yvette, France \\ $^{(5)}$Institut Non-Lin\'eaire de Nice,
Universit\'e de Nice-Sophia Antipolis, 1361 route des Lucioles, 06560
Valbonne, France} 

\date{\today}

\abst{
We present results from quantum Monte Carlo simulations of trapped bosons in
optical lattices, focusing on the crossover from a
gas of softcore bosons to a Tonks-Girardeau gas in a one-dimensional 
optical lattice. 
We find that depending on the quantity being measured, the behavior found in the 
Tonks-Girardeau regime is observed already at relatively small values of the interaction strength.
A finite critical value for entering the Tonks-Girardeau regime does not exist.
Furthermore, we discuss the computational efficiency
of two quantum Monte Carlo methods to simulate large scale trapped
bosonic systems: directed loops in stochastic series expansions and the
worm algorithm.}

\kword{boson Hubbard model, quantum Monte Carlo, cold atoms, optical
lattice, hard-core bosons, Tonks-Girardeau gas} 
 
\begin{document}
\maketitle

\section{Introduction}

Interest in the properties, phase transitions and various phases of
strongly correlated systems in reduced dimensionality has a very long
history in condensed matter physics. The experimental realization of
Bose-Einstein condensates (BEC) in traps~\cite{bec} offered the possibility
of studying such phenomena in precisely engineered systems. This goal
was eventually achieved by placing the condensates on optical
lattices~\cite{greiner,stoeferle} which has the effect of packing the bosonic
atoms in close proximity to one another and even producing multiple
occupancy of lattice sites. Consequently, strongly correlated systems
are produced on such optical lattices which have the advantage of
being defect free. However, the lattices are immersed in the confining
trap and are consequently not translationally invariant. This makes
delicate the task of applying results known in condensed matter
physics for translationally invariant systems to BEC on optical
lattices. The rush to do this has led to some misunderstandings
especially concerning ``quantum phase transitions'' on the optical
lattice.

Our goal here is to use quantum Monte Carlo (QMC) methods, such as the
stochastic series expansion (SSE) and the worm algorithm, to study in
some detail the properties of BEC on optical lattices and elucidate
the nature of the observed ``transition''.

The paper is organized as follows. In section~\ref{sec:model}, we
introduce the bosonic Hubbard model used for the description of a cold
atomic Bose gas trapped on optical lattices and discuss and compare
the techniques to simulate it. We then review the results presented in
Ref.~\citen{PRA} in section \ref{review}, focusing on observables particularly suited for the
detection of local ``phases'' or structures and the absence of quantum
criticality in these inhomogeneous systems. In
section~\ref{sec:tonks}, new results on the transition to a Tonks-Girardeau gas
regime for bosonic systems in one-dimension are presented.

\section{Quantum Monte Carlo simulations}
\label{sec:model}
\subsection{Model}
We consider the following Bose Hubbard Hamiltonian
\begin{eqnarray}
\label{eq:hamiltonian}
H&=&-t\sum_{\langle i, j \rangle} \left(b^{\dagger}_i b_j +
h.c.\right)\\ & & + \frac{U}{2}\sum_i n_i(n_i-1) + V \sum_i r^2_i n_i
- \mu \sum_i n_i,\nonumber
\end{eqnarray}
as a good description of the low energy physics of cold confined
bosonic gases~\cite{zoller}.  Here $t$ is the nearest neighbor
hopping, $U$ the onsite repulsion between the softcore bosons, $\mu$
the chemical potential and $V$ the curvature of the parabolic
confining potential imposed by the trap. We use standard notations for
the bosonic operators.  We furthermore define $\mu_i^{\rm
eff}=\mu-Vr_i^2$, which is the local effective chemical potential
experienced by a boson at site~$i$.
\subsection{Quantum Monte Carlo}
The Hamiltonian (\ref{eq:hamiltonian}) is particularly well suited for
simulations using QMC techniques. Indeed and to
our best knowledge, such techniques are the only ones able to simulate
efficiently the Hamiltonian (\ref{eq:hamiltonian}) for the following
reasons :

\begin{itemize}
\item They work in {\it any dimension} (unlike the density matrix renormalization group (DMRG) which is
restricted to one-dimensional systems).
\item They can be used to obtain both finite {\it and} zero temperature properties.
\item They allow the simulation of  {\it non homogeneous} trapping potentials
(unlike series expansion).
\item {\it Large system sizes} can be reached (unlike using exact diagonalization).
\item They perform an {\it exact treatment of all terms} in the
 Hamiltonian (\ref{eq:hamiltonian}) (unlike most analytic approaches).
\end{itemize}

\subsection{Efficiency of non-local quantum Monte Carlo\\ methods}
\label{sec:QMC}

We used the most recent world-line QMC algorithms, 
the SSE algorithm \cite{SSE} with directed loops
\cite{DirectedLoops1,DirectedLoops2} and the worm algorithm
\cite{WormCode} to perform the simulations based on the open source
implementations of the ALPS project.\cite{alps} 

The SSE representation \cite{SSE} starts from a Taylor expansion of
the partition function in orders of the inverse temperature $\beta$: 
\begin{eqnarray}
\label{eq:sse}
Z&=&{\rm Tr}\exp(-\beta H)=\sum_{n=0}^\infty\frac{\beta^n}{n!}{\rm
Tr}(-H)^n \\
&=&\sum_{n=0}^\infty\frac{\beta^n}{n!}\sum_{\{i_1,...i_n\}}\langle
i_1|-H|i_2\rangle
\cdots\langle i_n|-H|i_1\rangle . \nonumber
\end{eqnarray}
This representation of the partition function does not suffer from
errors introduced by time discretization and the sampling algorithm
can be combined with non-local updates such as directed loops
\cite{DirectedLoops1,DirectedLoops2}. However, the SSE representation
of the partition function corresponds to a perturbation expansion in
both, diagonal and off-diagonal, terms of the Hamiltonian. For the
simulation of trapped bosonic systems we are naturally interested in
the limit where the diagonal matrix elements (e.g. the trapping
potential) are large. In this limit the SSE sampling can face a
considerable slowdown as the probability of exchanging off-diagonal
and diagonal bond terms in the Hamiltonian is suppressed.  On the
contrary, the worm algorithm \cite{WormCode} samples world lines in
the path integral representation of the partition function which
treats only the off-diagonal terms in the Hamiltonian as a perturbation
\begin{eqnarray}
Z & = & {\rm Tr}\left( e^{-\beta H} \right) = {\rm Tr}\left( e^{-\beta
H_0} {\rm T} e^{-\int_0^{\beta} d\tau V(\tau)} \right)\;. \nonumber \\
\end{eqnarray}

The directed loop and worm algorithms both work in an extended
configuration space, which  in addition to closed world line 
configurations, allows for the presence of an open world line fragment, the ``worm" or
``directed loop", which is formally introduced by adding a source term
to the Hamiltonian
\begin{equation}
H_{\rm worm}=H-\eta\sum_i(b_i^++b_i)\;.  
\end{equation}
This source term allows the world lines to be broken with a matrix
element proportional to $\eta$. To generate new closed loop
configurations in the sampling process, a worm is created and a random
sequence of local updates of the worm is performed where each move
fulfills detailed balance. Nevertheless, this update procedure can
perform non-local changes of the world line configurations as the worm
can wind around the lattice in the temporal or spatial direction and
thereby change the particle or winding number respectively.
\begin{figure}
\begin{center}
\includegraphics[width=8cm]{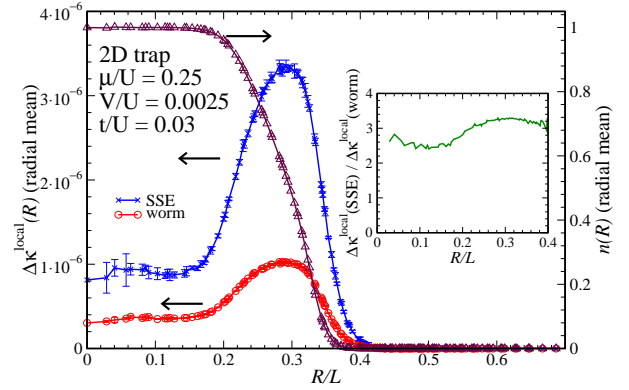}
\caption{Comparison of the statistical error $\Delta\kappa^{\rm local}$  of the
local compressibility $\kappa^{\rm local}$ of a 2D trap with $35 \times 35$ sites
obtained from simulation runs with fixed CPU-time for stochastic
series expansion (SSE) quantum Monte Carlo and the worm algorithm. The
error bars are obtained by averaging over estimates with identical
radial distance. The inset shows the ratio of statistical errors
$\Delta\kappa^{\rm local}({\rm SSE}) / \Delta\kappa^{\rm local}({\rm worm}) $ for the two
algorithms using the open-source implementations of the ALPS
project.\cite{alps}}
\label{fig:delta_kappa}
\end{center}
\end{figure}

To discuss the computational
performance of the two QMC algorithms in  simulating
confined ultra-cold bosonic atoms in two dimensions we
compare the statistical error obtained for the Monte Carlo estimates
of observables from simulation runs with a fixed amount of CPU time
(typically around 100 CPU hours on an Opteron 1.8 GHz processor).

While in the homogenous Bose-Hubbard Hamiltonian without confining
potential the SSE representation is found to perform better,  except from 
the extreme softcore case,\cite{Daniel} the situation is different when simulating
harmonic traps.  In Fig.~\ref{fig:delta_kappa} results of our
simulations are shown for a two-dimensional system with $35 \times 35$ sites and trap
parameters which allow for a Mott plateau with integer density ($\langle n_i \rangle = 1$) 
in the center of the trap ($R/L \lesssim
0.2$).
Before discussing the physics of this system in the next sections we first
consider the statistical error of the local compressibility $\kappa^{\rm local}$ (defined by Eq.~(\ref{eq:kappalocal}), below),
which is shown in Fig.~\ref{fig:delta_kappa} as a function of the distance from the trap center. While
for both algorithms the statistical error is largest for the
superfluid ring surrounding the central Mott plateau ($0.2 \lesssim R
\lesssim 0.4$), the worm algorithm yields a statistical error which is a factor of 3
smaller than the one obtained for the SSE
estimate with the same amount of CPU-time. 
The superior performance of
the worm algorithm is a consequence of the small ratio of off-diagonal
to diagonal matrix elements for the used value of $t/U = 0.03$ in the simulated Bose-Hubbard
Hamiltonian. Further reducing this ratio, e.g. to study trapped
bosonic systems which exhibit more than one Mott plateau, will result
in an even larger speedup of the worm algorithm.

\begin{figure}
  \begin{center} \includegraphics[width=8cm]{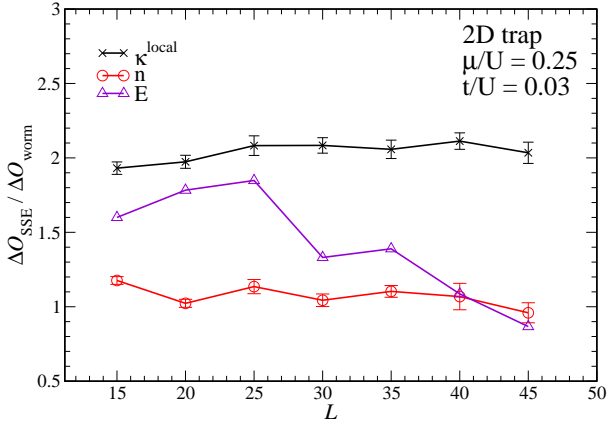}
  \caption{Scaling of the speedup of the worm algorithm compared to
  the SSE algorithm versus system size for the same trap parameters as
  in Fig.~\ref{fig:delta_kappa}. Shown are results for the spatially averaged local density $n$ and  
  compressibility $\kappa^{\rm local}$ as well as  the energy $E$. For all simulations 
  the open-source implementations of the ALPS project~\cite{alps} were used.}
  \label{fig:scaling} \end{center}
\end{figure}

Fig.~\ref{fig:scaling} shows the ratio of statistical errors of QMC
estimates for three different observables obtained using the SSE and worm
algorithm. The statistical errors are shown  versus the number of lattice sites.  
For the particle number $n$ and the compressibility $\kappa^{\rm local}$ (c.f. Eq.~(\ref{eq:kappalocal}))
spatial averages 
are shown. There appears to be a constant improvement of
the worm code over the SSE algorithm, independent of the 
system size, implying that both algorithms exhibit the same scaling in 
system size.  From our simulations we conclude that the worm
algorithm is the superior QMC technique to simulate
trapped ultra-cold bosonic systems.

\section{Simulations of trapped bosonic atoms}
\label{review}
\subsection{Coexistence of phases and identification of local structures}
\label{sec:id}

As shown in various studies (see for example
Refs.~\citen{greiner,stoeferle} for experimental and Refs.~\citen{zoller,batrouni}
for theoretical/numerical investigations), the inhomogeneity in the
system due to the trapping potential induces a co-existence of
superfluid and Mott insulator-like regions. This can be understood on qualitative grounds by looking
at density profiles: a region of space with constant integer density of particles is 
interpreted as a Mott insulating
region (a Mott plateau), while non-integer densities correspond to superfluid regions. To distinguish the state near a given 
site more precisely, we proposed a more quantitative probe, which will
be reviewed here\cite{PRA}.

\begin{figure}
\begin{center}
\vspace{0.5cm}
\includegraphics[width=8cm]{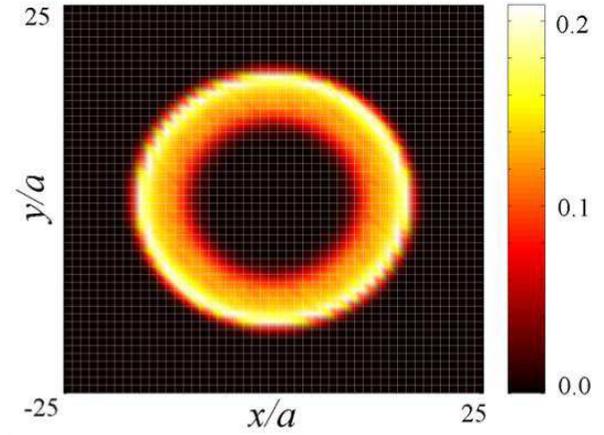}
\caption{Spatial dependence of the local compressibility $\kappa^{\rm
local}$, of bosons in a two dimensional parabolic trap with curvature
$V/U=0.002$, for $\mu/U=0.37$ and $U/t=25$.  A superfluid ring
surrounding a $n=1$ central Mott plateau is clearly resolved.
}
\label{fig:kappalocal}
\end{center}
\end{figure}

For the homogeneous case, the Mott insulating phase has a 
vanishing compressibility $\kappa$~\cite{fisher}. One way of
locally characterizing a region of space in the inhomogeneous system is
to study the local compressibility at a given lattice-site $i$:
\begin{equation}
\label{eq:kappalocal}
\kappa^{\rm local}_i=\left\langle\frac{\partial n}{\partial
\mu^{\rm eff}_i}\right\rangle=\beta\left(\langle
n_i n \rangle - \langle n_i \rangle \langle n \rangle \right),
\end{equation}
which expresses  the {\it total} density response of the
system to a {\it local} change of the chemical potential at site
$i$. This local compressibility
directly corresponds to the total compressibility $\kappa$ in the
homogeneous case, and serves to distinguish the local states in the inhomogeneous system:
$\kappa^{\rm local}_i$ is zero in a Mott insulating region while it remains finite
superfluid region. In Fig.~\ref{fig:kappalocal},
we show the spatial distribution of $\kappa^{\rm local}_i$ in
a two dimensional trapped system. Regions of space with vanishing $\kappa^{\rm
local}_i$ are clearly identified and correspond to Mott
insulating regions in the sample.

For the Hamiltonian in Eq.~(\ref{eq:hamiltonian}), the kinetic term $t$, the
repulsion $U$ and the trap curvature $V$ are tunable parameters in
the experiments,\cite{greiner,stoeferle,nature,moritz,tolra,kinoshita} and a ``quantum
phase transition'' between a Mott
insulator and a superfluid can be observed by changing e.g. the value of $U/t$. 
To study this transition, we calculated the momentum distribution function
\begin{equation}
\label{eq::greensfunction}
n({\bf k})=\frac{1}{N}\sum_{i,j} e^{i({\bf r}_i-{\bf r}_j){\bf k}}
\langle b^\dagger_i b_j \rangle,
\end{equation}
where $N$ is the total number of particles in the system.  The
momentum distribution in Eq.~(\ref{eq::greensfunction})
is normalized, and the coherence
fraction is given by the height of the coherence peak, $n({\bf
k}=0)$. Not only is this quantity useful to detect the
formation of a Mott plateau as will be shown below: it also is a 
quantity of interest for experiments, as $n({\bf k})$ can be determined 
from interference patterns~\cite{proko}.

\begin{figure}
\begin{center}
\vspace{0.5cm}
\includegraphics[width=8cm]{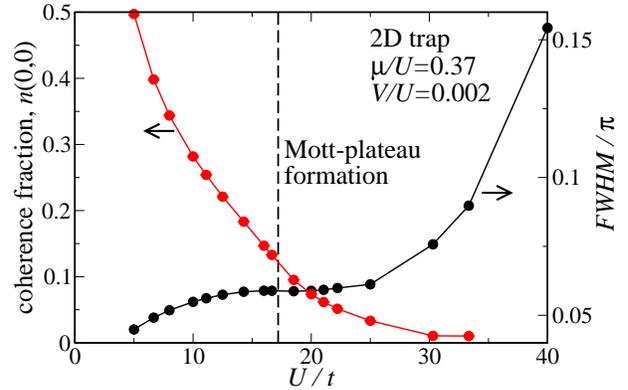}
\caption{ Evolution of the coherence fraction (the height of the peak,
$n(0,0)$) and of the full width at half maximum ($FWHM$) of the
coherence peak as a function of $U/t$ for bosons in a two dimensional
parabolic trap with curvature $V/U=0.002$ and with $\mu/U=0.37$. The
threshold for Mott plateau formation is indicated by the dashed line.
}
\label{fig:fwhm}
\end{center}
\end{figure}

Fig.~\ref{fig:fwhm} displays both the coherence
fraction $n({\bf k}=0)$ and the full width at half maximum (FWHM) of
the coherence peak in a two dimensional trap as obtained from the 
QMC simulations. 
The ratio $U/t$ at which a Mott plateau forms in the
center of the trap is marked by a vertical line. Whereas the coherence fraction does not show any
specific feature at the threshold for Mott plateau formation, we
observe a change of the curvature in the FWHM at this point. 
It was shown~\cite{PRA}, that this appears as a generic feature and
that the FWHM can thus serve as an indicator of Mott plateau formation in a confined Bose gas. 
We also showed that
the appearance of fine structure in the momentum distribution
(secondary peaks) is not related to the Mott plateau formation,
disproving claims in previous work~\cite{proko}.

\subsection{Absence of quantum criticality}
\label{sec:abs}

Fig.~\ref{fig:kappalocal} suggests that the superfluid region
identified by a non vanishing local compressibility is confined to
a shell around the central Mott-insulating region. 
One might then expect that in a two-dimensional trap this shell 
essentially behaves like a one-dimensional bosonic chain, and displays
similar quantum critical behavior. For example, the
compressibility of a one-dimensional bosonic chain diverges at the
superfluid-insulator transition~\cite{batrouni_old}. However,
simulations in a trap indicate no critical features in quantities like 
the local compressibility~\cite{PRA}.

To further investigate this phenomenon and to simplify simulations, an
effective bosonic ladder model was introduced in Ref.~\citen{PRA}, that
represents the superfluid ring in the original two-dimensional trap. 
In this model, each leg of the ladder is assigned a different chemical 
potential, which is chosen
such that the first leg is always in a $\left< n
\right> =0$ Mott insulating phase and the last leg in the $\left< n
\right> =1$ Mott insulating phase. The chemical potential is then
interpolated (for example linearly) between these two values for the
other legs of the ladder.

\begin{figure}
\begin{center}
\vspace{0.5cm} \includegraphics[width=8cm]{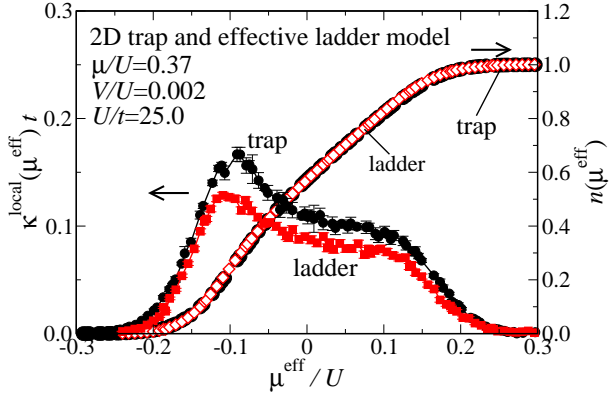}
\caption[]{ Local density, $n$, and local compressibility,
$\kappa^{\rm local}$, for a two dimensional parabolic trap on a square
lattice and for a ladder, as functions of $\mu^{\rm eff}$.  The
parameters for the trapping potential are $V/U=0.002$, at
$\mu/U=0.37$, and the parameters of the ladder model are chosen as to
cover the whole superfluid region (see Ref.~\citen{PRA}).}
\label{fig:ladder1}
\end{center}
\end{figure}

In Fig.~\ref{fig:ladder1}, we show the local density and the local
compressibility $\kappa^{\rm local}$ as a function of the effective
chemical potential $\mu^{\rm eff}$ for both simulations, the 2d trap
and the ladder model. The overall good agreement of the two sets of
curves indicates that indeed the ladder model captures the essential physics of
the trapped two-dimensional system. The density curves from
the trap and the ladder model coincide almost perfectly, 
indicating the validity of a local density approximation~\cite{PRA}. 
There are small differences in the local
compressibility between the trap and the ladder system, which are
due to the different shape of the local chemical potential
in the two systems. Nevertheless, the two different compressibility
curves share the same overall shape.

\begin{figure}
\begin{center}
\vspace{0.5cm}
\includegraphics[width=8cm]{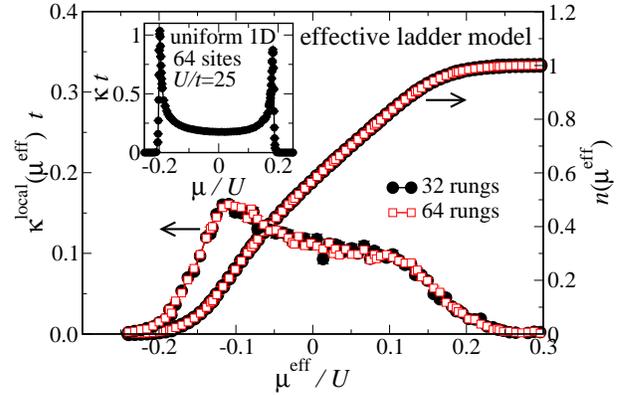}
\caption[]{ Local density, $n$, and local compressibility,
$\kappa^{\rm local}$, for the Bose-Hubbard model on ladders with
different lengths as a function of $\mu^{\rm eff}$.  The other
parameters of the ladder model are chosen as in
Fig.~\ref{fig:ladder1}.  The inset shows the compressibility for the
uniform one-dimensional case as a function of $\mu$ for a chain with
64 sites, for the same value of $U/t$ as used in the ladder model.}
\label{fig:ladder2}
\end{center}
\end{figure}

Having shown that the ladder model describes correctly the different
physical features found in the trapped system, we can  study the occurrence of quantum criticality in
this simpler model. In Fig.~\ref{fig:ladder2}, the local density and the local
compressibility of the ladder model are shown for two different system sizes. We
observe two very broad peaks in the compressibility, which might indicate quantum critical behavior. However, these peaks 
do not sharpen upon increasing the length of the ladder, as seen in Fig.~\ref{fig:ladder2}.
The inset of Fig.~\ref{fig:ladder2} displays the compressibility of a
one-dimensional chain which, in contrast with the ladder results, 
displays very sharp peaks at the insulating-superfluid phase transition,
even for a system of moderate size ($L=64$).

We thus find that no one-dimensional quantum critical behavior is found in the
realistically  simulated trapped systems. This is in agreement with the
absence of ``critical slowing down'' observed in the
experiments~\cite{greiner,stoeferle}. We thus  interpret the ``quantum phase
transition'' seen in the experiments as a crossover of changing
fractions of the Mott-insulating and superfluid regions. Critical
behavior could however be observed for flat confining
potentials. 
For a more elaborate discussion on the ladder model and
the absence of quantum criticality, we refer the reader to
Ref.~\citen{PRA}.

\section{Approaching the Tonks-Girardeau regime}
\label{sec:tonks}
\subsection{The Tonks-Girardeau gas}
In previous sections, we studied properties of confined ultra-cold
bosonic atoms in optical lattices using the Bose-Hubbard
Hamiltonian with an on-site repulsion $U/t$. In the
limit of infinite repulsion, $U/t\rightarrow \infty$, this model maps
onto a model of hardcore bosons which are constrained to a maximum
occupancy of one boson per site. In the continuum limit of the
one-dimensional Bose gas, the regime of infinite repulsion defines the
Tonks-Girardeau gas~\cite{tonks1,tonks2}, which in many respects
behaves similar to fermions due to the enforced impenetrability. The realization of
a Tonks-Girardeau gas in trapped bosonic atom systems~\cite{olshanii} initially
turned out difficult, due to the restricted interaction range
accessible from the experimental setup~\cite{moritz,tolra}.  However,
recently Tonks-Girardeau gas behavior was observed in one-dimensional
Bose gases in the presence of an optical lattice
potential~\cite{nature}. The underlying lattice structure increases
the effective mass of the bosons, thus increasing the ratio
$\gamma=E_{\rm int}/E_{\rm kin}$, of interaction energy ($E_{\rm int}$) to kinetic
energy ($E_{\rm kin}$)~\cite{nature}. In particular, excellent agreement
between the experimental data and calculations based on the
fermionization approach to the Tonks-Girardeau gas was obtained in
Ref.~\citen{nature}, already for values of $\gamma\approx 5$. In
another recent study~\cite{kinoshita}, observation of 
Tonks-Girardeau gas behavior was reported for  similar value of
$\gamma\approx 5$ also in the absence of an optical lattice.

\subsection{When do we observe Tonks-Girardeau gas behavior?}
In the following, we compare both the ground state density distribution and the momentum
distribution function of bosons in one-dimensional optical lattices
with a harmonic confinement potential to address the question when a system of repulsively interacting bosons
shows behavior similar to a Tonks-Girardeau gas.  We performed simulations for
chains of $N_s=50$ lattice sites, typical for experimental
setup~\cite{nature}, and considered different fillings of the system by
adjusting the chemical potential.

An appropriate means of quantifying the relevance
of the interactions in the presence of an optical lattice is 
the ratio $U/t$, given in terms of the microscopic model
parameters~\cite{cazalilla}. To a given value of $U/t$, an
associated effective value of $\gamma_{\rm eff}=U/(2t\langle n \rangle)$ is
obtained by determining the particle density $\langle n\rangle$
during the simulation~\cite{delft}.

\begin{figure}
\begin{center}
\vspace{0.5cm}
\includegraphics[width=8cm]{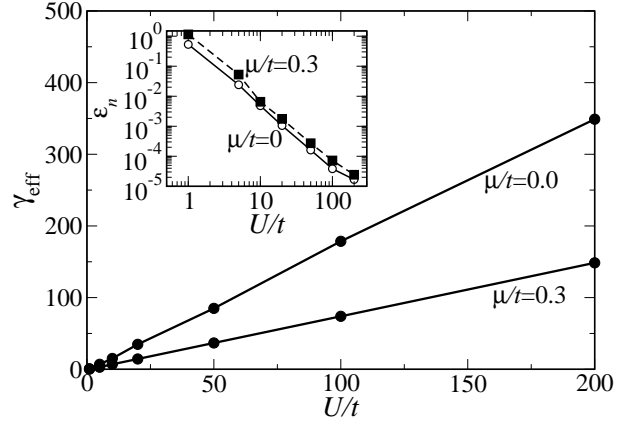}
\caption{
Effective interaction strength $\gamma_{\rm eff}$ as a
function of the ratio $U/t$ for bosons on one-dimensional optical
lattices , inside a harmonic confinement potential of strength
$V/t=10^{-3}$.  Results of quantum Monte Carlo simulations are shown
for different values of the chemical potential, $\mu/t=0$,
and $0.3$.  The inset shows the average deviation $\epsilon_n$
in the local density between the softcore and hardcore cases as a
function of the interaction strength $U/t$ for $\mu/t=0$ and $0.3$.  }
\label{fig:gamma}
\end{center}
\end{figure}

In Fig.~\ref{fig:gamma} the QMC results for $\gamma_{\rm eff}$ are shown as a function of $U/t$ for $\mu/t=0$ and $0.3$. 
For a given value of $U/t$, the effective value of the interactions, $\gamma_{\rm eff}$,
is larger for  $\mu/t=0$ than for $\mu/t=0.3$, due to a lower bosonic density in the first case.
We thus expect the behavior of softcore bosons to fit well to that of a Tonks-Girardeau gas
for smaller values of $U/t$ for $\mu/t=0$ than for $\mu/t=0.3$. 

\begin{figure}
\begin{center}
\vspace{0.5cm}
\includegraphics[width=8cm]{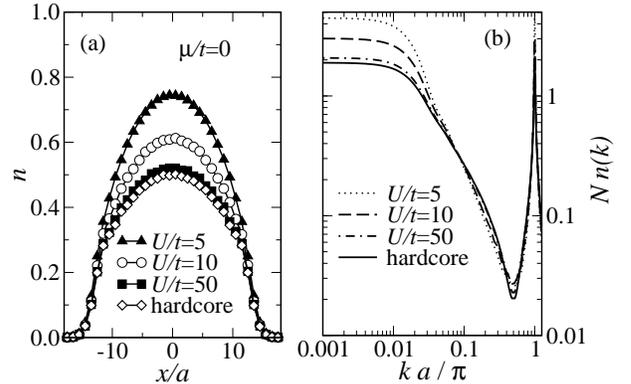}
\caption{
Density distribution, $n(x)$, and momentum distribution function,
$n(k)$, in the ground state of softcore bosons on one-dimensional
optical lattices for different values of $U/t=5$, $10$, $50$, and for
hardcore bosons ($U/t=\infty$), inside a harmonic confinement
potential of strength $V/t=10^{-3}$ and for $\mu/t=0$.}
\label{fig:softhard1}
\end{center}
\end{figure}

\begin{figure}
\begin{center}
\vspace{0.5cm}
\includegraphics[width=8cm]{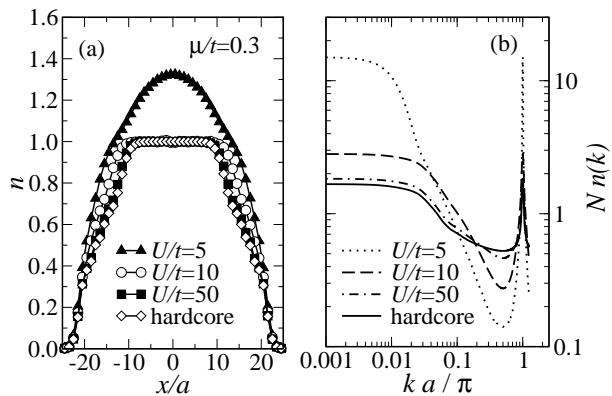}
\caption{
Density distribution, $n(x)$, and momentum distribution function,
$n(k)$, in the ground state of softcore bosons on one-dimensional
optical lattices for different values of $U/t=5$, $10$, $50$, and for
hardcore bosons ($U/t=\infty$), inside a harmonic confinement
potential of strength $V/t=10^{-3}$ and for $\mu/t=0.3$.}
\label{fig:softhard2}
\end{center}
\end{figure}

In Fig.~\ref{fig:softhard1} (a,b) the spatial density
distribution , $n(x)$, and the momentum distribution function, $n(k)$,
are shown in the low-density regime for $\mu/t=0$, with no
Mott-insulating region present in the trap.
We find that for
$U/t=50$ both quantities are already close to the hardcore limit.
Data taken at $U/t=100$ (not shown) did not exhibit any visible
difference to the hardcore limit in this regime.  In
contrast, for $U=5$ the density in the trap center is significantly
larger than in the hardcore case, as is the coherent fraction
$n(k=0)$.  One might thus conclude, as in Ref.~\citen{pollet}, that 
large values of $\gamma>100$ are needed before the behavior of
softcore bosons fits well to that of a Tonks-Girardeau gas.
Observing that near the boundary of the system, the
density distributions is close in all cases and that due to the low density
of bosons in this regime, the value of $U$ is irrelevant, 
we prefer a different look at this problem.

For a more detailed analysis, we plot the averaged squared deviation
$\epsilon_n(U)=1/N_s\sum_i (n_i(U)-n_i(U=\infty))^2$ as a function of
$U/t$ in the inset of Fig.~\ref{fig:gamma}. We find, that $\epsilon_n(U)$
decreases algebraically upon increasing $U$, i.e. $\epsilon_n(U)\propto U^{-\alpha}$,
with $\alpha\approx 2$. This shows that the crossover of softcore bosons 
into the hardcore limit is smooth, and that no finite critical value for $U/t$ exists.

Considering the momentum distribution function $n(k)$ shown in Fig.~\ref{fig:softhard1} (b), we find
that they are similar to the hardcore limit for all values of $U/t$  in this intermediate-density regime.
The main differences occur for the plateau value near $k=0$, whereas the characteristic slopes of the curves around
$k/a\approx \pi/2$~\cite{nature} are insensitive to the value of $U/t$. Due to this similar structure, and since
the overall scale of $n(k)$ is not fixed by the experimental data, 
the measured momentum profiles could be fitted in Ref.~\citen{nature} to 
the momentum profile of hardcore bosons already for moderate values of $\gamma$.
Experimentally it is thus difficult to distinguish the softcore regime from the hardcore limit
in the low-density region, and behavior similar to a Tonks-Girardeau gas emerges already for moderate 
values of $\gamma>5$, as observed in Ref.~\citen{nature}.
 
Next, we consider the case of an increased bosonic density inside the trap, obtained by increasing the
chemical potential. In particular, we consider $\mu/t=0.3$, and show results for the local density and
momentum distribution function in Fig.~\ref{fig:softhard2}. As seen from Fig.~\ref{fig:softhard2} (a), 
hardcore bosons develop an extended Mott plateau in the trap center for this value of $\mu$.  
Plateau formation is observed for softcore bosons only for values of $U/t>10$.
Furthermore, the algebraic decrease in $\epsilon_n(U)$ upon increasing $U$ found for $\mu/t=0$ is 
also observed for $\mu/t=0.3$ and with the same exponent of
$\alpha\approx 2$.
Turning to the momentum distribution functions,
we find from
Fig.~\ref{fig:softhard2} (b), that the momentum profile 
for $U/t=5$ and $10$ now show significant quantitative differences to the one in the hardcore limit.
In particular, the characteristic slopes of the curves near
$k/a\approx \pi/2$ now differ significantly from  the hardcore case. 
For high bosonic densities, the differences between
softcore and hardcore behavior are thus more pronounced, and easier detectable by 
measuring the experimentally accessible momentum profile.

\section{Conclusion}

The main new results of the present paper concern the crossover of softcore bosons into the Tonks-Girardeau regime. From our simulations we conclude 
that this
crossover appears
smooth, and that no finite critical value of the interaction strength $U/t$ exists.
Instead, upon increasing $U/t$, differences in the local density distributions between the softcore and 
hardcore case decrease algebraically, with an exponent that is insensitive to the density inside the trap. 

In the low-density region, the shape momentum profiles of softcore bosons are qualitatively very similar to
those observed in the Tonks-Girardeau gas already for moderate values of $U/t$. 
This explains the  observation in Ref.~\citen{nature}, 
where the measured momentum profiles could 
well be fitted using a fermionization approach to the Tonks-Girardeau gas for  
$\gamma>5$. 
The value of $\gamma$ or $U/t$ beyond which an experiment 
is well described by a Tonks-Girardeau gas picture
depends sensitively on the quantity being measured, the measurement error and the density of bosons. 
In particular in the low-density region, we find the momentum profile to be rather
insensitive to the value of $U/t$, allowing a description of the softcore boson system
in terms of a Tonks-Girardeau gas already at small values of $\gamma$.
For larger densities, 
the additional confinement of the superfluid in the hardcore limit 
leads to quantitative  differences in the momentum profile between
softcore and hardcore bosons and thus larger values of $U/t$ are needed 
in order to observe the behavior of a Tonks-Girardeau gas.

\section*{Acknowledgments}

The numerical calculations were performed using the SSE application package
with generalized directed loop techniques~\cite{DirectedLoops2} and
the worm code application package of the ALPS project,~\cite{alps} and
carried out on the Asgard and Hreidar Beowulf clusters at ETH Z\"urich.

We thank M.A.~Cazalilla, P. Denteneer, T.~Esslinger, A.~Muramatsu, L. Pollet, N.V.~Prokof'ev, M.~Rigol, and
P.~Zoller for fruitful discussions.

We acknowledge support from the Swiss National Science Foundation, the
Kavli Institute for Theoretical Physics in Santa Barbara and the Aspen
Center for Physics.

\end{document}